\documentclass[eqsecnum,twocolumn,aps]{revtex4}
\usepackage[english]{babel}
\usepackage{amssymb}
\usepackage{graphicx}

\begin{document}

\title{Nonlinear Spectroscopic Effects in Quantum Gases\\Induced by Atom--Atom Interactions}
\author{A.\,I.\,Safonov$^{a,b}$}\email{alesaf2008@rambler.ru}
\author{I.\,I.\,Safonova$^a$}
\author{I.\,S.\,Yasnikov$^c$}
\affiliation{$^a$National Research Centre Kurchatov Institute, 123182 Moscow, Russia\\
$^b$Moscow Institute of Physics and Technology (State University), 141700 Dolgoprudny, Moscow region, Russia\\
$^c$Togliatti State University, 445667 Togliatti, Russia}

\date{August 28, 2012; in final form, October 23, 2012}

\begin{abstract}We consider nonlinear spectroscopic effects -- interaction-enhanced double resonance and spectrum instability $-$ that
appear in ultracold quantum gases owing to collisional frequency shift of atomic transitions and, consequently, due to
the dependence of the frequencies on the population of various internal states of the particles. Special emphasis is
put to two simplest cases, (a) the gas of two-level atoms and (b) double resonance in a gas of three-level bosons, in
which the probe transition frequency remains constant.\end{abstract}

\maketitle

\section{Introduction}
As is well-known, interaction of a multilevel quantum system simultaneously with several resonance fields is, under
certain conditions, accompanied by nonlinear spectroscopic phenomena, coherent population trapping (CPN)~\cite{Agapiev}
and electromagnetically induced transparency (EIT)~\cite{EIT}, which are caused by the formation of a special ``dark''
superposition state immune to the resonance fields. These effects, however, are still linear in a sense that the
resonance transition frequencies in the quantum system are independent of the population of various states. In this
work, we consider another, generally speaking, a more general class of phenomena induced by exactly such a dependence
exemplified by collisional or contact frequency shift of intra-atomic (e.g., hyperfine) transitions in quantum gases
owing to the interaction between of the gas particles. In our previous work \cite{Saf_EPJD}, we showed that interaction
of a gas of three-level atoms with two light fields results, due to the contact shift, in a specific kind of double
resonance, interaction-enhanced double resonance (INEDOR). In addition, as will be shown below, a gas of two-level
atoms with a nonzero contact shift can exhibit spectrum instability of the resonance transition, namely, the dependence
of the resonance line shape and the final population of the levels on sweep direction, as well as on the relation
between the amplitude of the probe field, sweep rate and the magnitude of the contact shift.

The behavior of an arbitrary two-level system is commonly described in terms of effective spin
(pseudo-spin) $s=1/2$. We will also follow this representation, in each case attributing spin to a
particular pair of quantum states coupled by a resonance transition. In this work, we disregard
spin waves and related effects associated with spatial transport of spin polarization, which
actually corresponds to zero spin diffusion constant. Below we will discuss to what extent this
assumption corresponds to real physical systems. Spatial inhomogeneity is included only in the
analysis of a possible INEDOR spectrum line shape in a linear gradient of the external field.

\section{Nonlinear Dynamics of a Three-Level System}\label{sec:Nonlin3D}
In general, the evolution of a quantum system interacting with resonance fields is described by Liouville--von Neuman
equation for the components of the spin density matrix $\rho$ \cite{Neuman}
\begin{equation}\label{eq:LfN}
i\hbar\frac{\partial\rho}{\partial t} = [\hat{H}(t),\rho] - i\hat\mathcal{L}(t)\rho,
\end{equation}
where $\hat{H}(t)=\hat{H}_0+\hat{U}(t)$ is the Hamiltonian including the unperturbed term $\hat{H}_0$ and the
time-dependent perturbation $\hat{U}(t)$ due to the external ac field, $\hat\mathcal{L}$ is the Lindblad superoperator
\cite{Lindblad} responsible for dissipation and square brackets, as usually, denote quantum-mechanical commutation.

To obtain the general evolution equation for the components of the density matrix of a gas of three-level atoms
interacting with two resonance fields we use the previously derived relation for the contact shift of the transition
$|1\rangle\rightarrow|2\rangle$ in a spatially homogeneous gas in the presence of the third state
$|3\rangle$~\cite{Saf_JLTP}. We will be interested in coherent population of the states. In this case, the frequency
shift $\Delta_{ij}(\rho)\equiv\omega_{ij}(\rho)-\omega_{ij}(0)$ $(i,j=1,2,3)$ of the transition between the states
$|i\rangle$ and $|j\rangle$ at a certain nonzero gas density $n\equiv\textrm{Tr}(\rho_{ij})$ vanishes for fermions,
whereas for bosons in the absence of a Bose$-$Einstein condensate it is
\begin{eqnarray}\label{eq:Dom123rho}
\hbar\Delta_{12}&=&(\rho_{11}+\rho_{22})\delta\lambda_{12}+\nonumber\\
&+&(\rho_{22}-\rho_{11})\Delta\lambda_{12}+2\rho_{33}(\lambda^+_{23}-\lambda^+_{13}),\label{eq:Dom123rho_12}\\
\hbar\Delta_{13}&=&(\rho_{11}+\rho_{33})\delta\lambda_{13}+\nonumber\\
&+&(\rho_{33}-\rho_{11})\Delta\lambda_{13}+2\rho_{22}(\lambda^+_{23}-\lambda^+_{12}).\label{eq:Dom123rho_13}
\end{eqnarray}
Here, $\rho_{ij}$ are the components of the density matrix in terms of the eigen wavefunctions of the unperturbed
Hamiltonian $\hat{H}_0$, $\delta\lambda_{ij}=\lambda_{jj}-\lambda_{ii}$,
$\Delta\lambda_{ij}=\lambda_{ii}+\lambda_{jj}-2\lambda^+_{ij}$, $\lambda^\pm_{ij}\equiv\langle
ij|\lambda|ij\rangle_\pm$ is the spin part of the matrix element of the interaction intensity $\lambda=4\pi\hbar^2a/m$,
which is commonly used to describe cold collisions, when the partial amplitudes of scattering with a nonzero angular
momentum of the relative motion of the colliding particles ``freeze out'', $m$ is the atomic mass, $a$ is the
respective s-wave scattering length. The superscript ``+'' denotes that the wavefunction of two colliding bosons is
symmetric with respect to permutation of both their pseudo-spin and spatial coordinates. The doubly antisymmetric
component, obviously, do not contribute to s-wave scattering.

\begin{figure}
\resizebox{\columnwidth}{!}{\includegraphics{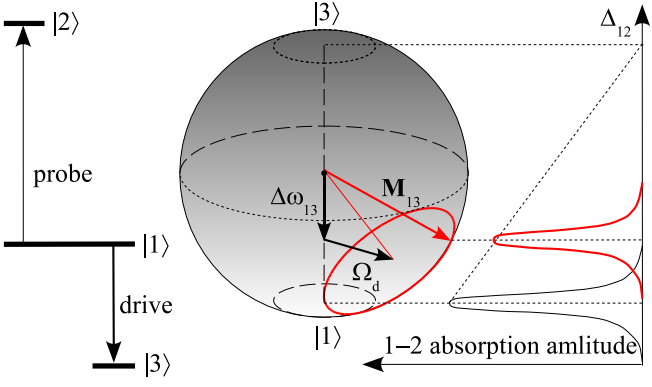}} \caption{Scheme of (left) a three-level system and (right)
influence of the Rabi oscillations between the states $|1\rangle$ and $|3\rangle$ on the $|1\rangle-|2\rangle$
transition frequency and intensity. For clarity, the Rabi period $2\pi\Omega_{\rm d}^{-1}$ of the drive transition is
assumed to be much longer than the detection time of the $|1\rangle-|2\rangle$ resonance line.} \label{fig:Bloch}
\end{figure}

Below we restrict ourselves to the typical situation in double-resonance experiments, when the state $|2\rangle$ is
initially unpopulated and its population changes insignificantly during the interaction with a weak probe field with
the frequency $\omega_{\rm p}$, whereas the populations of the states $|1\rangle$ and $|3\rangle$ can vary quite
arbitrarily under the action of a strong drive field with the frequency $\omega_{\rm d}$. In this case,
$\rho_{22}\ll\rho_{11},\rho_{33}$, and therefore the last term in the right-hand side of Eq.~(\ref{eq:Dom123rho_13})
can be omitted. Thus, the general equation (\ref{eq:LfN}) for the evolution of the components of the density matrix of
the gas of three-level bosons with a ladder ($\Xi$) level scheme (Fig.~\ref{fig:Bloch}) in the absence of spontaneous
longitudinal relaxation becomes (cp.~\cite{Agapiev})
\begin{eqnarray}\label{eq:SDM}
i\frac{\partial{\rho}_{11}}{\partial t}&=&\frac{\Omega_{\rm d}}{2}\rho_{13}\exp\left[i\Delta\omega_{13}(\rho)t\right]-\nonumber\\
&-&\frac{\Omega_{\rm p}}{2}\rho_{21}\exp\left[i\Delta\omega_{12}(\rho)t\right]-\textrm{c.c.},\label{eq:SDM_11}\\
i\frac{\partial{\rho}_{33}}{\partial t}&=&-\frac{\Omega_{\rm d}}{2}\rho_{13}\exp\left[i\Delta\omega_{13}(\rho)t\right]+\textrm{c.c.},\label{eq:SDM_33}\\
i\frac{\partial{\rho}_{31}}{\partial t}&=&\frac{\Omega_{\rm d}}{2}(\rho_{33}-\rho_{11})\exp\left[i\Delta\omega_{13}(\rho)t\right]+\nonumber\\
&+&\frac{\Omega_{\rm p}}{2}\rho_{32}\exp\left[i\Delta\omega_{12}(\rho)t\right]-i\gamma_{13}\rho_{31},\label{eq:SDM_31}\\
i\frac{\partial{\rho}_{22}}{\partial t}&=&\frac{\Omega_{\rm p}}{2}\rho_{21}\exp\left[i\Delta\omega_{12}(\rho)t\right]-\textrm{c.c.},\label{eq:SDM_22}\\
i\frac{\partial{\rho}_{12}}{\partial t}&=&\frac{\Omega_{\rm p}}{2}(\rho_{22}-\rho_{11})\exp\left[i\Delta\omega_{12}(\rho)t\right]+\nonumber\\
&+&\frac{\Omega_{\rm d}}{2}\rho_{32}\exp\left[i\Delta\omega_{13}(\rho)t\right]-i\gamma_{12}\rho_{12},\label{eq:SDM_12}\\
i\frac{\partial{\rho}_{32}}{\partial t}&=&-\frac{\Omega_{\rm d}}{2}\rho_{12}\exp\left[i\Delta\omega_{13}(\rho)t\right]-\nonumber\\
&-&\frac{\Omega_{\rm p}}{2}\rho_{31}\exp\left[i\Delta\omega_{12}(\rho)t\right]-i\gamma_{32}\rho_{32},\label{eq:SDM_32}
\end{eqnarray}
where $\Omega_{\rm p(d)}$ and $\Delta\omega_{12(13)}(\rho)\equiv\omega_{\rm p(d)}-\omega_{12(13)}(\rho)$ are the Rabi
frequency and density-dependent frequency detuning of the probe (drive) field and $\gamma_{ij}$ are the transverse
relaxation rates. The light fields are thought to be spatially homogeneous, which usually holds for hyperfine
transitions, whose wavelengths are much larger that the geometrical size of the sample.

If, as already mentioned above, the probe field is much weaker than the drive field, $\Omega_{\rm p}\ll\Omega_{\rm d}$,
the second term in Eqs. (\ref{eq:SDM_11}) and (\ref{eq:SDM_31}) can be neglected. In this case, Eqs.
(\ref{eq:SDM_11})--(\ref{eq:SDM_31}) can be written in a more compact form of a usual Bloch equation for the precession
of the spin polarization (``magnetization'') vector ${\bf M}$ in the Hilbert space of the states $|1\rangle$ and
$|3\rangle$ (${\bf M}_z=(0,0,\rho_{33}-\rho_{11})$, ${\bf M}_\perp=2({\rm Re}\rho_{13},{\rm Im}\rho_{13},0)$):
\begin{equation}\label{eq:Bloch_31}
\frac{\partial{\bf M}}{\partial t} = {\bf M}\times\left(\tilde{\bf\Omega}+\frac{\Delta\lambda_{13}}{\hbar}{\bf
M}_z\right)-\gamma_{13}{\bf M}_\perp,
\end{equation}
where $\tilde{\bf\Omega}=(\Omega_{\rm d},0,\omega_{13}(0)-\omega_{\rm
d}+\hbar^{-1}\delta\lambda_{13}(\rho_{11}+\rho_{33}))$. For clarity, we separate in the precession frequency
$\tilde{\bf\Omega}$ the component that remains nearly constant in a weak probe field, when
$\rho_{11}+\rho_{33}\approx$~const. The second term in the parenthesis in the right-hand side of Eq.
(\ref{eq:Bloch_31}) expresses explicitly the dependence of the precession frequency on the current value of the
magnetization vector {\bf M}. In the subsequent sections, we discuss the direct consequences of this, generally
speaking, nonlinear precession in two limiting cases.

\section{Interaction-Enhanced Double Resonance}\label{sec:INEDOR}
As follows from Eq. (\ref{eq:Dom123rho_12}), excitation of the transition from the state
$|1\rangle$ to $|3\rangle$ in a Bose gas induces frequency modulation of the transition
$|1\rangle-|2\rangle$ associated with the Rabi oscillations of the populations of the states
$|1\rangle$ and $|3\rangle$~\cite{Saf_JLTP}. If both transitions are excited simultaneously, this
modulation leads to a new phenomenon, interaction-enhanced double resonance.

Here, for simplicity and clarity, we restrict ourselves to the case $\lambda_{11}=\lambda^+_{12}$, when the contact
shift of the transition $|1\rangle-|2\rangle$ vanishes at $\rho_{22}=\rho_{33}=0$. Another advantage of this case is
that it allows direct comparison with the experiments on electron-nuclear double resonance (ENDOR) in atomic
hydrogen~\cite{Ahokas,Ahokas_JLTP}. To simplify our consideration even further, we assume that the frequency of the
drive transition $|1\rangle-|3\rangle$ remains constant, which requires $\Delta\lambda_{13}$ to be sufficiently small.
An opposite case is considered in Sec.~\ref{sec:Instability}. Despite the above assumptions, a general analytic
solution of nonlinear system (\ref{eq:SDM_22})--(\ref{eq:Bloch_31}) is hardly possible. The approach implemented below
does not feature such generality but provides quantitative and physically transparent description in the case of
interest.

Physics of the novel kind of double resonance becomes clear from Fig.~\ref{fig:Bloch}. In the absence of relaxation,
the populations of the states $|1\rangle$ and $|3\rangle$ oscillate under continuous excitation of the transition
$|1\rangle-|3\rangle$ in the initially pure state $|1\rangle$ sample at the frequency (see Fig.~\ref{fig:Bloch})
$\tilde{\Omega}_{13}=\sqrt{\Omega_{\rm d}^2+\Delta\omega_{13}^2}$. This results in the frequency modulation of the
probe transition $|1\rangle-|2\rangle$~\cite{Saf_JLTP}
\begin{equation}\label{eq:bc_dyn}
\hbar\omega_{12}=\hbar\omega_{12}(0)+2n\Delta\lambda\left(\frac{\Omega_{\rm
d}}{\tilde{\Omega}_{13}}\right)^2\sin^2\left(\frac{\tilde{\Omega}_{13}t}{2}\right),
\end{equation}
where $\Delta\lambda=\lambda^+_{23}-\lambda^+_{13}$. The amplitude of this frequency modulation can easily be
comparable with or even much greater than the linewidth of the probe transition. In this case, excitation of the
transition $|1\rangle-|3\rangle$ periodically drives the probe transition out of resonance. For clarity,
Fig.~\ref{fig:Bloch} corresponds to slow driving in a sense that the Rabi period $2\pi\Omega_{\rm d}^{-1}$ of the drive
transition is thought to be much longer than the time needed to detect the $|1\rangle-|2\rangle$ resonance line and the
entire $|1\rangle-|2\rangle$ spectrum moves periodically fore and back along the frequency axis and simultaneously
changes in amplitude. The waveforms shown in Fig.~\ref{fig:Bloch} are the schematic ``snapshots'' of the spectrum at
different phases of the Rabi cycle. Obviously, electromagnetic absorption at the fixed frequency $\omega_{\rm p}$ of
the probe field also changes periodically with a period of the $|1\rangle-|3\rangle$ Rabi oscillations. It should be
emphasized that these changes are associated with changes in \textit{both} the population of the initial state
\textit{and} the transition frequency, in contrast to conventional double resonance, which is solely due to a change in
the population of the initial state caused by the drive transitions. Clearly, such interaction-induced frequency
modulation can greatly enhance the effect, which therefore can be called INteraction-Enhanced DOuble Resonance
(INEDOR).

\begin{figure*}
\resizebox{1.01\columnwidth}{!}{\includegraphics{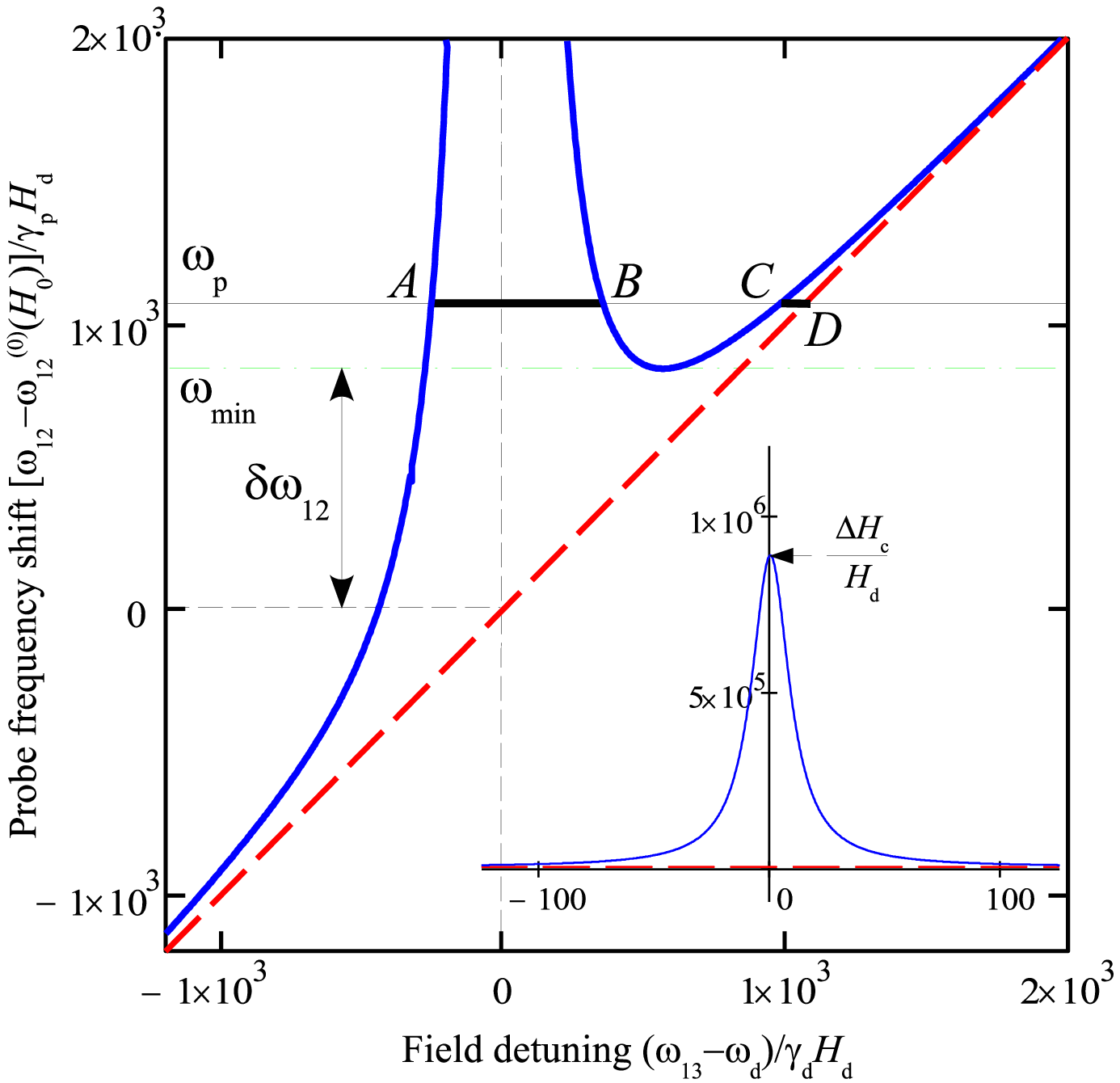}}\resizebox{0.99\columnwidth}{!}{\includegraphics{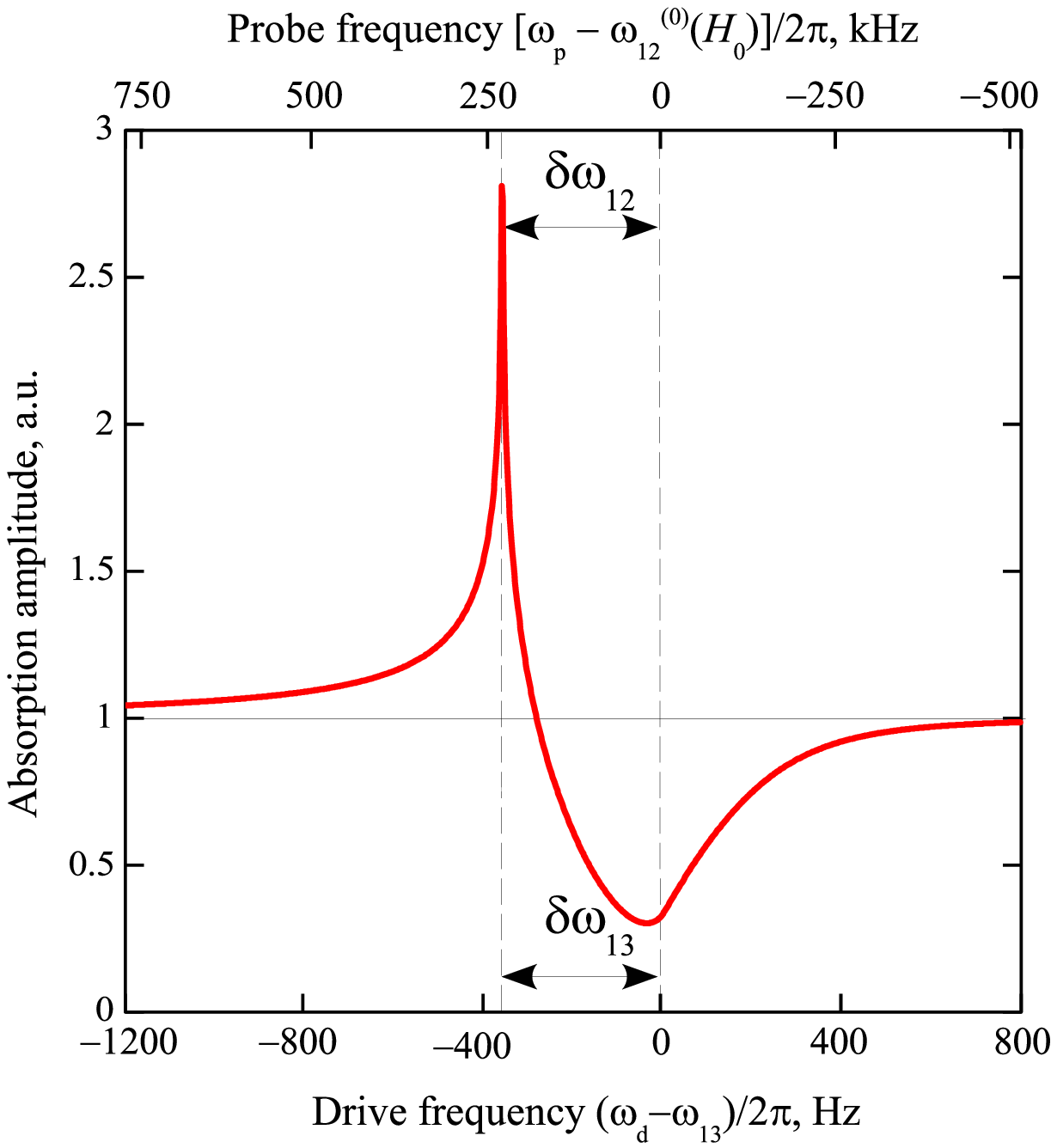}}
\caption{(a) Field dependence of the $|1\rangle-|2\rangle$ frequency shift (in units of $\gamma_{\rm p}H_{\rm d}$)
under the CW excitation of the $|1\rangle-|3\rangle$ resonance. Horizontal axis is the field detuning
$(\omega_{13}-\omega_{\rm d})/\gamma_{\rm d}$ from the $|1\rangle-|3\rangle$ resonance in units of the excitation field
$B_{\rm d}$. Solid and dashed line are, respectively, the upper bound of the sum of the Zeeman and mean-field
contributions and the Zeeman contribution alone. Horizontal lines indicate (dash-dotted line) the value of
$\omega_{12}$ at minimum and (solid line) the probe frequency. Vertical dashed line corresponds to the resonance field
for the $|1\rangle-|3\rangle$ transition. Inset is the overview of the same dependence. (b) $|1\rangle-|2\rangle$
absorption amplitude as a function of the drive frequency (the lower horizontal axis is the detuning, in Hertz, from
the $|1\rangle-|3\rangle$ resonance) for the monochromatic probe field with the fixed frequency $\omega_{\rm
p}=\omega_{12}(0,B_0)$. Sharp peak corresponds to the minimum of the probe frequency in (a). The spectrum that appears
when $\omega_{\rm p}$ (upper horizontal axis) is swept at fixed $\omega_{\rm d}=\omega_{13}$ is inverted with respect
to the frequency axis. The parameters correspond to 2D atomic hydrogen with a density of $3\cdot10^{12}$ cm$^{-2}$ in
the high polarizing field $B=45$~kG, except for the sign of the contact shift: $B_{\rm d}=1$~mG, $\Delta B_{\rm
c}=89$~G.}\label{fig:INEDOR}
\end{figure*}

Let us consider in more detail a possible line shape of the INEDOR spectrum in a strong drive field when, in contrast
to the case illustrated in Fig.~\ref{fig:Bloch}, simultaneous frequency and amplitude modulation of the
$|1\rangle-|2\rangle$ absorption line is relatively fast and therefore is integrated by the detection system. This
implies that the $|1\rangle-|3\rangle$ Rabi frequency is much higher than the rate of field or frequency sweep through
the $|1\rangle-|2\rangle$ resonance or the inverse time constant $\tau^{-1}$ of the detection system, $\Omega_{\rm
d}\tau\gg1$.

Generally, the energies of all three levels and, consequently, both transition frequencies depend on the external
static field. The specific nature of this field is insignificant but for definiteness we shall consider the magnetic
field $B$. Let the field $B_0$ correspond to the exact resonance $|1\rangle-|3\rangle$ at a certain density $\rho$,
$\omega_{13}(\rho,B_0)=\omega_{\rm d}$. Then the deviation $b=B-B_0$ of the field from this value is equivalent to the
frequency detuning $\Delta\omega_{13}(\rho,b)=\gamma_{\rm d}b$ and $\Delta\omega_{12}(\rho,b)=\gamma_{\rm p}b$ of the
drive and probe field, respectively (here, $\gamma_{\rm d(p)}$ is the corresponding gyromagnetic ratio). Owing to this
Zeeman contribution to the frequency $\tilde{\Omega}_{13}(\rho,b)=\gamma_{\rm d}\sqrt{B_{\rm d}^2+b^2}$ of the
magnetization precession forced by the drive field $B_{\rm d}(t)=B_{\rm d}\exp(i\omega_{\rm d}t)$, the amplitude of the
oscillating component of $\omega_{12}$ in Eq.~(\ref{eq:bc_dyn}), i.e., the amplitude of the frequency modulation of the
probe transition, is a Lorentzian function of the static field $b$. As a result, $\omega_{12}(t)$ oscillates between
the Zeeman-only zero-density lower bound (dashed line in Fig.~\ref{fig:INEDOR}a)
\begin{equation}\label{eq:omega12_stat}
\omega_{12}(0,B_0+b)=\omega_{12}(0,B_0)+\gamma_{\rm p}b
\end{equation}
and the upper bound, which is the sum of the Zeeman and contact-shift contributions (thick solid line in
Fig.~\ref{fig:INEDOR}a),
\begin{equation}\label{eq:12_tot}
\omega_{12}(\rho,B_0+b)=\omega_{12}(0,B_0)+\gamma_{\rm p}b+\left(\frac{2n\Delta\lambda}{\hbar}\right)\frac{B_{\rm
d}^2}{B_{\rm d}^2+b^2},
\end{equation}
at the field-dependent frequency $\tilde{\Omega}_{13}(b)$~\cite{Saf_EPJD}. According to Eq.~(\ref{eq:12_tot}), the
upper bound of the probe frequency is generally a nonmonotonic function of the external static field.

The time-average absorption amplitude $A(b, \omega_{\rm p})$ at the probe frequency within the bounds
(\ref{eq:omega12_stat}) and (\ref{eq:12_tot}) is proportional to the average population of the initial state and the
probability density to find the system at given values of $b$ and $\omega_{\rm p}$. The probability density, in turn,
peaks at the lower and upper bound of the probe transition frequency, where the system spends more time, since in these
cases, $d\omega_{12}(b, \omega_{\rm p})/dt=0$ \cite{Saf_JLTP}.

To illustrate the effect of spatial inhomogeneity, let us consider in more detail an infinite sample with a spatially
homogeneous density in a linear gradient $\nabla B$ of the static field and homogeneous light fields. In this case,
different parts of the sample simultaneously experience all possible field values. The absorption amplitude at a given
probe frequency $\omega_{\rm p}$ is proportional to the integral $I(\omega_{\rm p})=\int A(b, \omega_{\rm
p})\left(\frac{\partial N}{\partial b}\right)db$ along the line $\omega_{\rm p}={\rm const}$ within the segments $AB$
and $CD$ in Fig.~\ref{fig:INEDOR}a. The result of numerical integration with the parameters of the ENDOR experiments
with 2D atomic hydrogen at a density of $3\cdot10^{12}$~cm$^{-2}$ in the strong polarizing magnetic field $B=45$~kG and
the drive field $B_{\rm d}=1$~mG~\cite{Ahokas,Ahokas_JLTP,Vas_priv}, at the maximum amplitude $\Delta B_{\rm
c}=2n\Delta\lambda(\hbar\gamma_{\rm p})^{-1}=89$~G  of the contact shift of the hyperfine transition
$|b\rangle-|c\rangle$ in field units (here, $\gamma_{\rm p}$ and $\gamma_{\rm d}$ are to a good accuracy equal to the
gyromagnetic ratio of electron and proton, respectively), is shown in Fig.~\ref{fig:INEDOR}b as a function of the drive
frequency $\omega_{\rm d}$ (lower horizontal axis) at constant $\omega_{\rm p}=\omega_{12}(0,B_0)$. Alternatively, the
INEDOR spectrum can be detected by sweeping $\omega_{\rm p}$ (upper horizontal axis) at constant $\omega_{\rm d}$ (in
Fig.~\ref{fig:INEDOR}b, $\omega_{\rm d}=\omega_{13}(0,B_0)$). In this case, the spectrum is inverted on the frequency
scale because an increase in $\omega_{\rm d}$ corresponds to an increase in the resonance value of the static field
$\omega_{\rm d}/\gamma_{\rm d}$ and, consequently, to a positive displacement of the mean-field Lorentzian peak
(\ref{eq:12_tot}) on the curve $\omega_{12}(b)$ in Fig.~\ref{fig:INEDOR}a. The latter is in turn equivalent to a
decrease in the probe frequency $\omega_{\rm p}$ at constant $\omega_{\rm d}$. Which way of observing the INEDOR
spectrum is preferred depends on the details of a particular experiment.

The absorption amplitude as a function of $\omega_{\rm d}$ (Fig.~\ref{fig:INEDOR}b) decreases substantially within the
drive resonance and has a sharp maximum at $\omega_{\rm p}=\omega_{\rm min}$, i.e., when the minimum value of
Eq.~(\ref{eq:12_tot}) coincides with the probe frequency. This explains the dispersion-looking ENDOR spectra of 2D
atomic hydrogen~\cite{Ahokas,Ahokas_JLTP}. Qualitatively, the hole in the absorption amplitude is because the atoms of
the otherwise resonant regions of the sample are periodically driven out of the probe resonance and, as a result, spend
only a small fraction of time in the resonance conditions. On the other hand, the absorption maximum is due to the fact
that the drive resonance introduces zero gradient of the $|1\rangle-|2\rangle$ transition frequency in a certain region
of the sample and, therefore, much more atoms become resonant.

The width of the double-resonance curve in the probe-frequency units can be estimated as the difference
$\delta\omega_{12}=\omega_{\rm min}-\omega_{12}(0,B_0)$ (see Fig.~\ref{fig:INEDOR}a). It is readily shown that for a
sufficiently high amplitude of the contact shift such that $|2n\Delta\lambda|\gg\hbar\gamma_{\rm p}B_{\rm d}$ and $b\gg
B_{\rm d}$~\cite{Saf_EPJD},
\begin{equation}\label{eq:width_w13}
\delta\omega_{13}=\frac{\gamma_{\rm d}}{\gamma_{\rm p}}\delta\omega_{12}\simeq\frac{3}{2}\gamma_{\rm d}(2\Delta B_{\rm
c}B^2_{\rm d})^{1/3},
\end{equation}
where $\Delta B_{\rm c}=2n\Delta\lambda(\hbar\gamma_{\rm p})^{-1}$ is the maximum contact shift of the
$|1\rangle-|2\rangle$ resonance in field units. Thus, the INEDOR linewidth is independent of the static field gradient.
On the other hand, the spectrum intensity is inversely proportional to $|\nabla B|$.

\section{Spectrum Nonlinearity of a Gas of Two-Level Bosons}\label{sec:Instability}
\begin{figure}
\resizebox{\columnwidth}{!}{\includegraphics{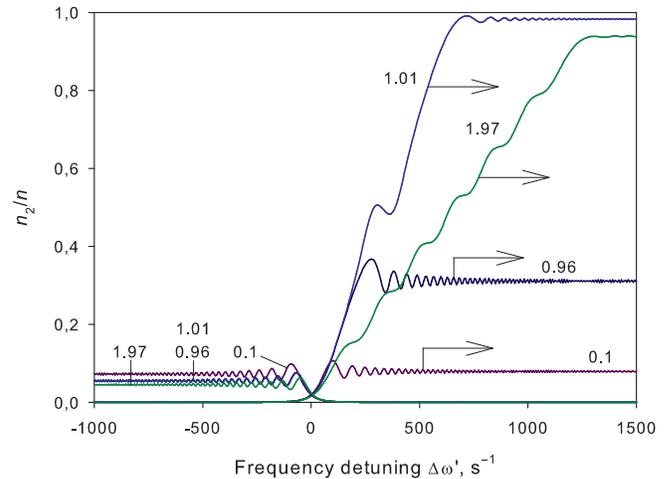}} \caption{Evolution of the population of the state $|3\rangle$
during the linear upward (indicated by the right arrow) and downward frequency sweep of the microwave field according
to the numeric solution of Eq.~(\ref{eq:Bloch_31}). The total density corresponding to each curve is marked in units of
the critical density $n_c$. The values of the parameters correspond to the hyperfine transition $b\rightarrow a$ in
three-dimensional atomic hydrogen: $\Omega=10$~s$^{-1}$, $d\omega/dt=2\times10^3$~s$^{-2}$,
$n_c\approx2.23\times10^{18}$~cm$^{-3}$, $\gamma=0.3$~s$^{-1}$, $\Gamma=0$\cite{Vas_priv},
$\Delta\lambda/\hbar=-3\times10^{-16}$~cm$^3$/s.}\label{fig:Instability-1}
\end{figure}

\begin{figure}
\resizebox{\columnwidth}{!}{\includegraphics{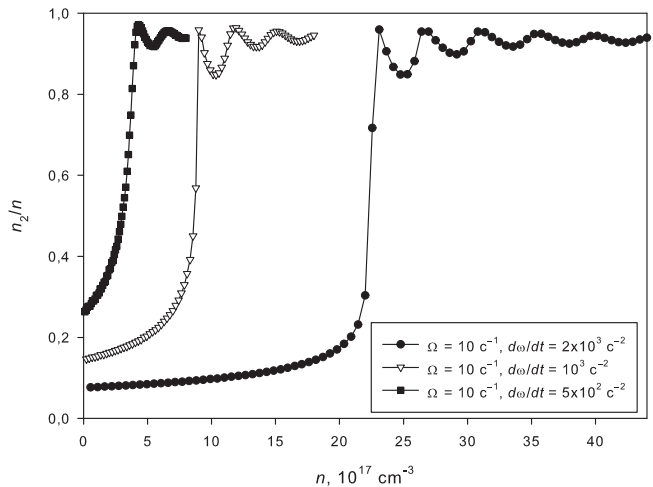}} \caption{Calculated total-density dependence of the final
population of the state $|3\rangle$ at various frequency sweep rates $|d\omega/dt|$. Other parameters are the same as
in Fig.~\ref{fig:Instability-1}.}\label{fig:Instability-2}
\end{figure}

In contrast to usual optical Bloch equations, Eq. (\ref{eq:Bloch_31}) includes an essentially nonlinear term
proportional to the contact shift of the resonance frequency. To observe this nonlinearity it is sufficient to have
just two levels. To avoid confusion, we keep levels 1 and 3 omitting for brevity the subscripts 13 and p(d). Thus, in
this section, $\Delta\lambda\equiv\Delta\lambda_{13}$. In addition, we set the total density constant,
$n\equiv\rho_{11}+\rho_{33}=\textrm{const}$. The nonlinearity is seen most clearly in the spectrum at a low enough
sweep rate. In particular, if the rate $\hbar^{-1}\Delta\lambda dM_z/dt\sim\hbar^{-1}\Omega n\Delta\lambda$ of the
variation of the transition frequency due to a change in the populations is greater than the frequency sweep rate
$d\omega/dt$, the Rabi oscillations occur on the background of a relatively slow change in the magnetization along with
the frequency of the light field, so that the system stays all the time near the resonance
$M_z\Delta\lambda\approx\hbar[\omega(\rho)-\omega(0)]-n\delta\lambda$. Meanwhile, the populations periodically vary
relatively quickly when the resonance conditions are fulfilled, which in turn drives the system out of resonance for a
certain time until the frequency of the light field is adjusted to the new value of the transition frequency. The
picture of such a reentrant resonance repeats until the transition frequency stops changing because the limiting value
of the magnetization (e.g., $M_z=n$) is reached. Frequency sweep in the opposite direction is accompanied by usual
behavior of the populations, since the repeated fulfillment of the resonance conditions is impossible. On the other
hand, it is clear that the contact frequency shift affects the spectrum considerably if it drives the system out of the
resonance conditions. This requires that the contact shift were at least comparable with the Rabi frequency (the
bandwidth of the microwave generator is assumed to be sufficiently narrow; for example, the spectral width of the
generator in the experiments with atomic hydrogen was $\delta f/f\lesssim10^{-10}$~\cite{Vas_priv}) Thus, under the
conditions
\begin{equation}\label{eq:contact instability}
\Omega n|\Delta\lambda|\gtrsim\hbar\left|\frac{d\omega}{dt}\right|; \;n|\Delta\lambda|\gtrsim\hbar\Omega
\end{equation}
there appears a spectrum ``hysteresis'' (Fig.~\ref{fig:Instability-1}). If in addition
$n|\Delta\lambda|\gg\hbar\Omega$, a drastic change in the spectrum upon reaching a certain critical value of the total
gas density $n_c$ or any other parameter entering the first condition of (\ref{eq:contact instability}) occurs almost
abruptly (Fig.~\ref{fig:Instability-2}).

Figure~\ref{fig:Instability-1} shows the evolution of the population of the state $|3\rangle$ during the upward and
downward (indicated by the arrows) linear frequency sweep of the microwave field, according to the numeric solution of
Eq.~(\ref{eq:Bloch_31}) with the parameters corresponding to the hyperfine transition $|b\rangle\rightarrow|a\rangle$
in three-dimensional atomic hydrogen, namely, at the Rabi frequency $\Omega=10$~s$^{-1}$, sweep rate
$d\omega/dt=2\times10^3$~s$^{-2}$ and the total density $n\sim2\cdot10^{18}$~cm$^{-3}$ \cite{Vas_priv},
$\Delta\lambda/\hbar=-3\times10^{-16}$~cm$^3$/s (see Appendix). The character of the spectrum is almost insensitive to
the transverse relaxation rate (see below). In the calculations, we used $\gamma=0.3$~s$^{-1}$. The longitudinal
relaxation rate in atomic hydrogen is quite low and was therefore neglected. The total gas density corresponding to
each curve is given in units of $n_c\approx2.23\times10^{18}$~cm$^{-3}$.

The calculated dependence of the final population of the state $|3\rangle$ on the total gas density at various sweep
rates $|d\omega/dt|$ and the same values of the other parameters as in Fig.~\ref{fig:Instability-1} is shown in
Fig.~\ref{fig:Instability-2}. When the density increases from a subcritical to supercritical value, the final
population of the state $|3\rangle$ after sweeping through the resonance conditions sharply increases from the low
value $\rho_{33}\sim\Omega^2|d\omega/dt|^{-1}$ given by the product of the Rabi frequency $\Omega$ and the duration
$\Omega|d\omega/dt|^{-1}$ of the sweep through the resonance line, to nearly unity. A further increase in density is
accompanied by a mere increase in the frequency detuning, at which the limiting population is reached. As is easily
seen, this detuning is exactly the maximum possible contact frequency shift $n\Delta\lambda$. At the opposite sweep
direction, the final population of the state $|3\rangle$, on the contrary, decreases with an increase in density. As
one might expect, the final population is independent on the sweep direction in the low-density limit, when the contact
shift vanishes. Thus, the shape of the spectrum depends on the rate and direction of the frequency sweep of the light
field, as well as on the field amplitude.

As is seen in Fig.~\ref{fig:Instability-2}, the critical density $n_c$ (determined as an abscissa of the steepest
slope) does not exactly conform to the condition $|\Omega n_c\Delta\lambda|\propto\hbar|d\omega/dt|$, which follows
from (\ref{eq:contact instability}), and rather increases superlinearly with $|d\omega/dt|$. The origin of this
behavior has to be clarifies. Note only, that the period $2\pi/\Omega$ of the Rabi oscillations at a low rate of the
microwave frequency sweep (left curve in Fig.~\ref{fig:Instability-2}) becomes comparable with the duration of passing
through the resonance line. Damping oscillations of the final population of the state $|3\rangle$ seen at $n>n_c$
presumably originate from the fact that the system finally gets out of the resonance conditions in this or that
particular phase of effective Rabi oscillations (waves on the slanted parts of curves in Fig.~\ref{fig:Instability-1}
at $n>n_c$).

Equation (\ref{eq:Bloch_31}) holds for a spatially homogeneous system. In the case of spatial inhomogeneity, there
appears magnetization transport owing to exchange and dipole--dipole interactions of the atomic pseudo-spins, which
leads to the emergence of spin waves \cite{Bashkin, Vainio}. The effect of the contact shift of the spin-wave spectrum
requires separate consideration, which lies beyond the scope of the present work.

The effect described in this section can occur not only in atomic hydrogen but also in ultracold alkali vapors and a
number of other systems. In any case, the particular character of interaction, which results in the dependence of the
transition frequency on the population of the states involved is insignificant. The reason why the spectrum
nonlinearity was not observed in the experiments on the contact shift in $^{87}$Rb vapor~\cite{Harber} is that the
second condition of (\ref{eq:contact instability}) was violated. In fact, the s-wave scattering lengths of $^{87}$Rb in
various hyperfine states are such that the maximum differential contact shift at the density $n\sim10^{13}$~cm$^{-3}$
was as small as $n\Delta\lambda/h\sim0.2$~Hz ($\Delta\omega/\omega\sim3\cdot10^{-11}$), whereas the Rabi frequency of
the two-photon transition in the pulsed microwave/RF field was about 2.5~kHz.

The dependence of the resonance field on the sample magnetization knowingly leads to a so-called ferromagnetic
instability~\cite{Anderson_Suhl}, e.g., of the FMR spectrum of ferromagnetic films, ESR in atomic
hydrogen~\cite{Instability} and NMR in $^3$He~\cite{Nacher}. In contrast to this type of instability, whose condition
is determined the transverse relaxation rate, the effect considered in this section does not depend directly on the
transverse relaxation, as the contact shift in a two-level system is independent of the mutual coherence of the
single-particle states~\cite{Zwierlein}.

We are truly grateful to S. A. Vasiliev for fruitful discussions and providing us with the data of
atomic hydrogen experiments the University of Turku, Finland. This work was supported by the Human
Capital Foundation, contract no. 211).

\appendix
\section{Contact Shift in Atomic Hydrogen} The magnitude of the contact shift of the hyperfine transition
$|b\rangle\leftrightarrow|a\rangle$ in atomic hydrogen can be found as follows. The diatomic states in the basis
$|S,m_S;I,m_I\rangle$ of the total electron and nuclear spins of the pair of atoms take the form
\begin{eqnarray}
|bb\rangle&=&|1,-1;1,-1\rangle\label{eq:H-H_2bb}\\
|ab\rangle_+&\equiv&\frac{1}{\sqrt{2}}\left(|ab\rangle+|ba\rangle\right)=\nonumber\\
&=&\cos\theta|1,-1;1,0\rangle-\sin\theta|1,0;1,-1\rangle,\label{eq:H-H_2ab}\\
|aa\rangle&=&\cos^2\theta|1,-1;1,1\rangle+\sin^2\theta|1,1;1,-1\rangle-\nonumber\\
&-&\frac{\sin2\theta}{2}|1,0;1,0\rangle-\frac{\sin2\theta}{2}|0,0;0,0\rangle\label{eq:H-H_2aa}
\end{eqnarray}
where $\tan(2\theta)=A[(\gamma_e+\gamma_p)hB]^{-1}$, $\gamma_e$($\gamma_p$) is the gyromagnetic ratio of electron
(proton), $A/h=1420$~MHz is the hyperfine constant of hydrogen). That is, the states $|bb\rangle$ and $|ab\rangle_+$
are pure electronic and nuclear triplets irrespective of the value of magnetic field. Consequently,
$\lambda^+_{ab}=\lambda_{bb}=\frac{4\pi\hbar^2}{m}a_t$, where $a_t$ is the triplet s-wave scattering length. Thus,
according to Eq.~(\ref{eq:Dom123rho_13}), the contact shift of the transition $|b\rangle-|a\rangle$ vanishes at
$\rho_{aa}=0$ in an arbitrary field. On the other had, the state $|aa\rangle$ contains the singlet component (the last
term in the hight-hand side of Eq. (\ref{eq:H-H_2aa})). Consequently,
$\Delta\lambda_{ab}=\lambda_{bb}+\lambda_{aa}-2\lambda^+_{ab}=\frac{\pi\hbar^2}{m}(a_s-a_t)\sin^22\theta\neq0$, так как
$a_t-a_s=30(5)$~pm \cite{Comment}. In the field $B=4.5$~T, $\Delta\lambda_{ab}/\hbar=-3\times10^{-16}$~cm$^3$/s, and
the frequency shift at the density $\rho_{aa}=2\cdot10^{18}$~cm$^{-3}$ is about $-100$~Hz, which is two orders of
magnitude greater than the Rabi frequency.

Replacing the electron spin by the nuclear spin all the aforesaid is automatically generalized to the transitions
$|b\rangle\leftrightarrow|c\rangle$, $|a\rangle\leftrightarrow|d\rangle$ and $|c\rangle\leftrightarrow|d\rangle$,
wherefrom $\Delta\lambda_{bc}=\Delta\lambda_{ad}=\Delta\lambda_{cd}=\frac{\pi\hbar^2}{m}(a_s-a_t)\sin^22\theta$.
However, the populations of the states with opposite electron spins cannot be simultaneously high owing to a fast
recombination of such pairs; thus, the sift of the transitions $|b\rangle\leftrightarrow|c\rangle$ and
$|a\rangle\leftrightarrow|d\rangle$ is hardly detectable.

\end{document}